 \documentclass[final,5p,times,twocolumn]{elsarticle}
\usepackage{amssymb}
\usepackage{xcolor}
\usepackage{comment}
\usepackage{braket}
\usepackage{mathtools}
\usepackage{gensymb}
\usepackage{multirow}
\usepackage[utf8]{inputenc}
\usepackage[T1]{fontenc}

\journal{Nuclear Instruments and Methods in Physics Research Section A}

\begin{document}

\begin{frontmatter}



\title{Depletion depth measurements of new large area silicon carbide detectors}

\author[lns]{A. Spatafora}
\author[lns]{D. Carbone}
\author[lns]{L. La Fauci}
\author[lns]{G. A. Brischetto}
\author[to]{D. Calvo}
\author[lns,unict]{F. Cappuzzello}
\author[lns]{M. Cavallaro}
\author[rbi]{A. Crnjac}
%\author[to]{M. Ferrero}
\author[rbi]{K. Ivanković Nizić}
\author[rbi]{M. Jakšić}
%\author[rbi]{G. Provatas}
\author[lns]{D. Torresi}
\author[lns]{S. Tudisco}

\author[]{for the NUMEN collaboration}

\affiliation[lns]{organization={INFN-Laboratori Nazionali del Sud},
             city={Catania},
             country={Italy}}
\affiliation[to]{organization={INFN-Sezione di Torino},
             city={Torino},
             country={Italy}}
\affiliation[rbi]{organization={Ruder Bošković Institute, Division of Experimental Physics},
             city={Zagreb},
             country={Croatia}}
\affiliation[unict]{organization={Dipartimento di Fisica e Astronomia "Ettore Maiorana", Università di Catania},
             city={Catania},
             country={Italy}}
\begin{comment}
\affiliation[unito]{organization={Università degli Studi di Torino, Dipartimento di Fisica},
             city={Torino},
            country={Italy}}

\affiliation[imm]{organization={ Institute for Microelectronics and Microsystems (IMM), National Research Council (CNR)},
             city={Catania},
             country={Italy}}    
\end{comment}

%%\affiliation{organization={},%Department and Organization

\begin{abstract}
The ion beam induced charge technique with proton microprobe is used to characterise newly developed p-n junction large area silicon carbide detectors. They were recently produced as part of the ongoing program to develop a new particle identification wall for the focal plane detector of the MAGNEX magnetic spectrometer at INFN - Laboratori Nazionali del Sud in view of the NUMEN experimental campaigns. Four silicon carbide devices are studied. Proton beams over a 1.26 to 6.00 MeV incident energy range are used to probe the active area and the depletion depth of each device. The energy loss tables for the silicon carbide material are checked, finding an empirical correction that is then used to quantify the depletion depth at the full depletion voltage through energy loss measurements of 3.40 MeV proton beams irradiating the back side of the devices. It is possible to fully deplete the devices provided that the epitaxial layer is grown properly on the substrate.  
\end{abstract}


\end{frontmatter}

\section{Introduction}
\label{sec:introduction}

Silicon Carbide (SiC) is a promising material for the next generation of particle detectors due to its high radiation hardness, energy and time resolution. It is the case of the PARIDE project \cite{paride}, which is planning to construct a new portable and versatile particle identification system made of SiC-Cesium Iodide (CsI) tallium doped (Tl) telescope detector cells, suitable to detect heavy ions in nuclear physics reactions. Also the NUMEN project \cite{numen2018, cappuzzello2021TDR} will use large area SiC detectors as the $\Delta$E stage of the new particle identification wall of the MAGNEX magnetic spectrometer \cite{magnex_review} at INFN-Laboratori Nazionali del Sud. 

State-of-the-art single pad SiC detectors were recently produced to these purposes \cite{tudisco2018}. They were built with 100 $\mu$m epitaxy, 10 $\mu$m substrate in the back side, 15.4 x 15.4 mm$^2$ area including 400 $\mu$m edge structure. Since the SiC detectors will be used to measure the energy loss of ions crossing them, the accurate knowledge of thickness and charge collection uniformity within the detectors are important requirements. For the same reason, it is important to study and validate the energy loss tables (e.g. Ziegler el al. \cite{Ziegler1985}) adopted for energy loss calculations in software like SRIM \cite{SRIM2010} or LISE++ \cite{Lise++}.  Moreover, within the NUMEN project, 720 SiC detectors will be used to cover the large MAGNEX detection area (154 x 1260 mm$^2$), thus homogeneity among different devices, in terms of depletion voltage, depletion depth, and resolution, is crucial. 

First characterizations of such SiC sensors were recently carried out \cite{carbone2024} by studying the I-V and C-V characteristics under irradiation with $\alpha$-source. To further test to what extent the mentioned requirements are fulfilled, we performed accurate studies at Ruđer Bošković Institute Accelerator Facility (RBI-AF) exploiting the ion beam induced charge (IBIC) microprobe technique \cite{angell1989,breese2007,vittone2013,manfredotti1995} with proton beams. This method is widely used for measuring the charge transport properties in solid state detectors. 

In this paper, we present and discuss a study of the energy loss for SiC material and the SiC detector depletion depth measurements. The paper is organized as follows. We describe the tested SiC sensors in Section \ref{sec:sic_properties} and the experimental setup in Section \ref{sec:exp_setup}. Then, the energy loss table validation is discussed in Section \ref{sec:validation}. Section \ref{sec:depletion_depths} presents the measurement of the depletion depth. Conclusions are drawn in Section \ref{sec:conclusions}.

\section{Large area Silicon Carbide detectors}
\label{sec:sic_properties}

SiC is a compound semiconductor with a wide bandgap ($\approx$ 3.28 eV), which significantly reduces the rate of thermal noise. Indeed, SiC devices present about 3 orders of magnitude smaller leakage current with respect to Silicon (Si) detectors \cite{Sze2006}. One of the most relevant characteristics is the radiation hardness, i.e. the resistance of the detectors to high doses of particle irradiation \cite{tudisco2025}. For example a dose as high as 10$^{13}$ ions/cm$^2$ of $^{16}$O ions at 25 MeV incident energy is sustainable in SiC compared to a deterioration of the signals starting from 10$^{9}$ ions/cm$^2$ in Si detectors \cite{altana2023}. These and other specific features of SiC detectors \cite{carbone2024, tudisco2018} make them a valuable alternative to Si detectors for applications in particle detection. 

\begin{figure}
   \centering
   \includegraphics[width=0.35\textwidth]{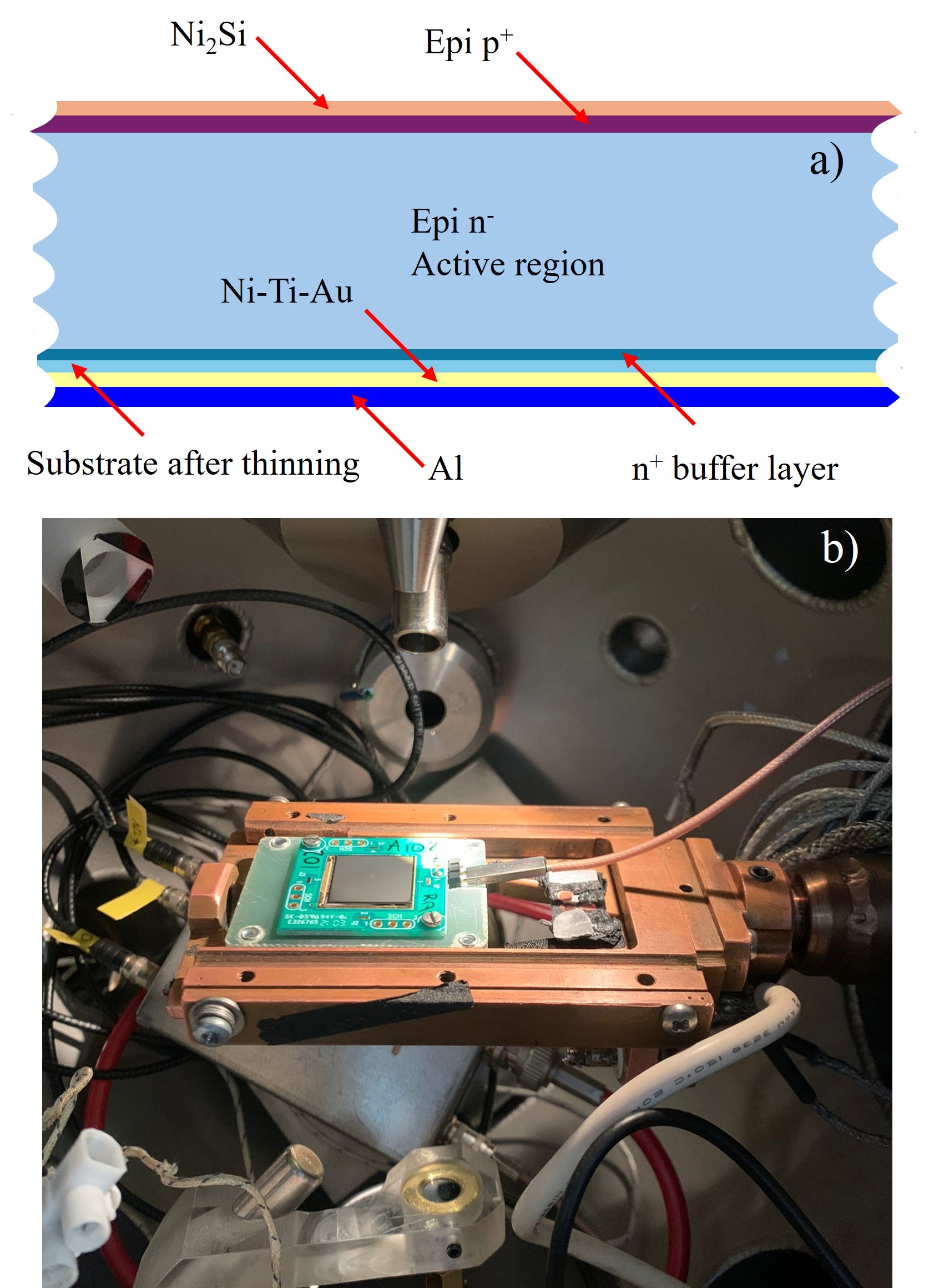}
    \caption{Views of the TT0012-11 A41 SiC device. a) Sketch of the sectional view illustrating the active area structure of the SiC device (not to scale). b) Picture of the device mounted on the PCB board inside the scattering chamber.\label{fig:sic_photoes}}
\end{figure}

A few prototype devices were constructed from two epitaxial wafers as described in Ref. \cite{carbone2024}. The different layers present in the active area of the SiC section are sketched in Figure \ref{fig:sic_photoes}a. They are produced with a 100 $\mu$m epi-layer grown on 350 $\mu$m SiC substrates. The device front size is metallized by a nickel silicide (Ni$_2$Si) deposition (thickness $\approx$ 100 nm). On the back side of the device, a mechanical thinning procedure is performed to reduce the total thickness of the device to $\approx$ 110 $\mu$m, which corresponds to a dead layer of $\approx$ 10 $\mu$m. Then, the ohmic contact is formed by an aluminum (Al) deposition ($\approx$ 1 $\mu$m). 

The 6" diameter epitaxial wafers used for producing the SiC detectors characterised in this work came from the same bulk material and were designated as "TT0012-11" and "RA0089-27". Different doping concentration was applied to the two wafers: the TT0012-11 wafer was doped with the standard concentration of $\approx$ 9 $\times$ 10$^{13}$ atoms/cm$^3$ corresponding to a Full Depletion Voltage (FDV) of $\approx$ 800 V, whereas the RA0089-27 wafer was doped with $\approx$ 3 $\times$ 10$^{13}$ atoms/cm$^3$ (FDV $\approx$ 350 V).  The device features a square area of 15.4 mm sideways. The total thickness as measured using a micrometer was 110 $\pm$ 1 $\mu$m. An edge structure $\approx$ 400 $\mu$m wide runs along the whole perimeter of the sensor. On the front, a small pad (0.15 $\times$ 0.30 mm$^2$) for the wire bonding is located in the centre of one side. 

First results on SiC detector performances were recently obtained studying the I-V and C-V characteristics and from $\alpha$-source irradiation tests \cite{carbone2024}. In general, the tested SiC sensors display a good energy resolution ($\approx$ 0.5 \% FWHM). Devices belonging to the TT0012-11 wafer displayed good C-V characteristics and doping profile. On the other hand, sensors belonging to the RA0089-27 wafer are characterised by worse performances especially in terms of the doping profile and the production yield. Preliminary estimations of the depletion depths were deduced using the energy loss tables of Ziegler et al. \cite{Ziegler1985}, embedded in the LISE++ code \cite{Lise++} and highlighted a smaller depletion than the expected $\approx$ 100 $\mu$m. These results stimulated the further tests discussed here.

In this work, the test performed on the A41 and A45 SiC sensors of the TT0012-11 wafer and A102 and A106 SiC sensors of the RA0089-27 wafer are reported. Their FDV, as obtained in Ref. \cite{carbone2024}, are listed in Table \ref{tab:deadlayer}.

\begin{table*}
    \centering
    \caption{Characteristics of the analysed SiC sensors: wafer name, sensor ID, FDV, dead layer thickness at the FDV evaluated from H microbeam and $\alpha$-source irradiation (from Ref. \cite{carbone2024}), and their average. The corresponding average depletion depth is also listed.}
\label{tab:deadlayer}
    \small
    \begin{tabular}{ccccccc} 
    \hline
    \multirow{2}{*}{Wafer} &\multirow{2}{*}{Sensor ID} & \multirow{2}{*}{FDV (V)} & \multicolumn{3}{c}{Dead Layer ($\mu$m)}& Depletion depth ($\mu$m)  \\
       & & & H beams & $\alpha$-source\textsuperscript{*}& average &average\\
%        &  & ($\mu$m) &  ($\mu$m) &  ($\mu$m) &  ($\mu$m) &  ($\mu$m) \\
    
    \hline
    TT0012-11 & A41 & 800 &  10.7  $\pm$ 0.3 & 10.2  $\pm$ 0.1  & 10.3  $\pm$ 0.1  & 99.8 $\pm$ 0.1\\
    TT0012-11 & A45 & 800   & 11.3  $\pm$ 0.1 & 11.4  $\pm$ 0.1 & 11.4  $\pm$ 0.1 & 98.7 $\pm$ 0.1\\
    RA0089-27 & A106 & 320 & 15.6  $\pm$ 0.1 & 15.3  $\pm$ 0.2 & 15.5  $\pm$ 0.1  & 94.5 $\pm$ 0.1\\
    RA0089-27 & A102 & 240 & 15.8  $\pm$ 0.2 & 16.1  $\pm$ 0.3 & 15.9  $\pm$ 0.2 & 94.1 $\pm$ 0.2\\
    \hline
    \end{tabular}
    \begin{flushleft}
    \textsuperscript{*} values from Ref. \cite{carbone2024} adopting the K normalization factor 
    \end{flushleft}
\end{table*}

\section{Experimental setup}
\label{sec:exp_setup}

The SiC devices were irradiated by a proton microbeam exploiting the IBIC technique implemented at RBI-AF. This technique is based on the use of MeV energy ion beams that create charge pairs in the detector's active region by ionisation \cite{Jaksic2022}. The SiC Sensors were mounted on a 30 x 30 mm$^2$ PCB board with a suitably sized opening of 15.2 x 15.2 mm$^2$. The board was attached to a mechanical actuator. It allows to position the detector with submillimetric accuracy and to vary the beam incidence angle on the device ensuring a precision of $\approx$ 0.1$^\circ$.

\begin{comment}
    accuratezze
    in X: each thick 0.01 inch = 0.254 mm
    in Y e Z: each thick 0.001 inch = 0.0254 mm = 25 um
\end{comment}

Regarding the pulse processing electronics, a 40 mV/MeV charge sensitive preamplifier, placed inside the vacuum chamber, was used to read out the SiC signal and to distribute the bias, supplied by an ORTEC 710 HV module. Then the signal was shaped and amplified by an ORTEC 572 amplifier after which it was digitized by Canberra ADC 8701 connected to the acquisition PC, equipped with the SPECTOR software \cite{cosic2019}. A similar electronic chain, based on ORTEC 142A preamplifier and ORTEC 570 amplifier, was adopted for a passivated implanted planar silicon (PIPS) detector of 300 $\mu$m thickness, used as the stop detector for some experimental runs. 

Several measurements were performed during the beam time, exploiting several different proton energies provided by the two tandem accelerators available at RBI-AF. Experimental runs with protons at 1.26, 2.50, 3.40, 3.92 MeV impinging on the front side of the SiC sensors were performed for energy calibration purposes.  To calibrate the PIPS detector the same proton energies and also a run a 6.00 MeV were used. Detectors were irradiated at the FDV in the central region where the charge collection efficiency is assumed to be 100\%. 

\begin{figure}
    \centering
    \includegraphics[width=0.65\linewidth]{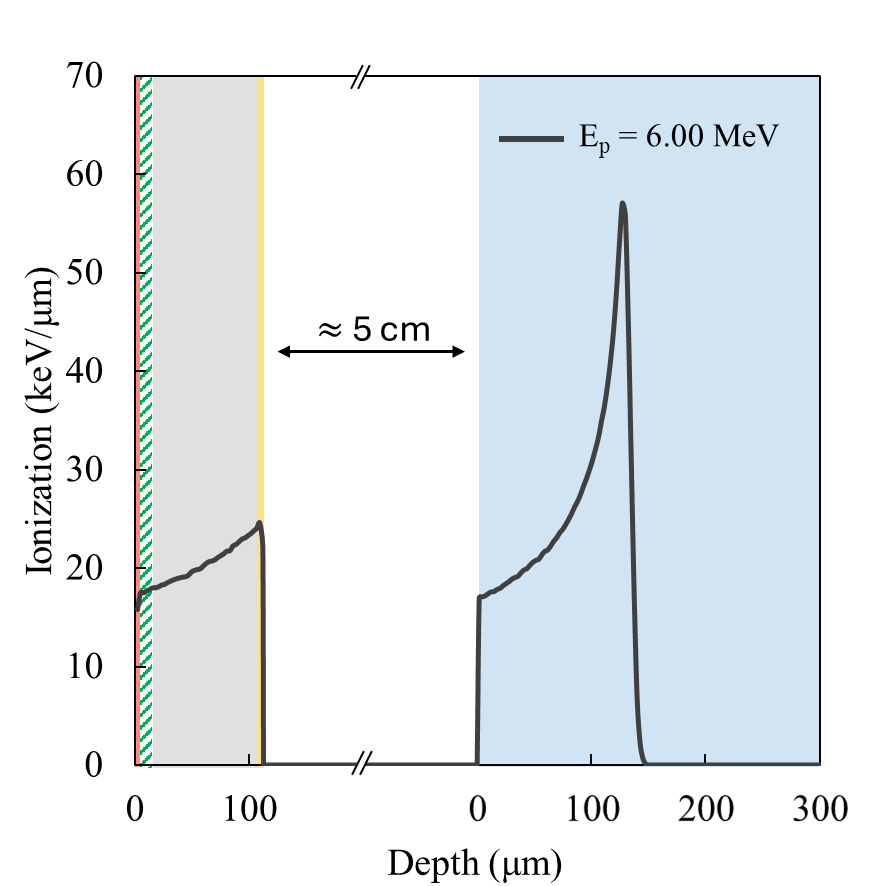}
    \caption{Ionization profile for 6.00 MeV proton energy, impinging at $\theta$ = 0$^\circ$, as obtained from simulations performed with the SRIM software \cite{SRIM2010} using the Ziegler energy loss tables \cite{Ziegler1985}. Different materials are considered corresponding to the different layers of the SiC and the PIPS detectors: Al (red band), SiC (gray band for the active region and green hatched band for the dead layer), Ni$_2$Si (yellow band), Si (blue band). The vacuum area of $\approx$ 5 cm between SiC and PIPS is indicated as the white band.}
    \label{fig:ranges}
\end{figure}

For the study of energy loss in SiC material, the SiC devices were irradiated on the back side in an inner area with 6.00 MeV protons at 6 different incident angles ($\theta$ = 0$^\circ$, 10$^\circ$, 30$^\circ$, 40$^\circ$, 50$^\circ$). The protons crossed the SiC detector and stopped in the PIPS one, which was placed downstream the SiC, used to measure the proton's residual energy. The ionization profile of the 6.00 MeV protons is reported in Figure \ref{fig:ranges}, as obtained from simulations performed with the SRIM software \cite{SRIM2010} using the Ziegler energy loss tables \cite{Ziegler1985}.
Experimental measurements in which the SiC detectors were irradiated from the back side with 3.40 MeV protons at different incident angles ($\theta$ = 0$^\circ$, 20$^\circ$, 30$^\circ$, 40$^\circ$, 50$^\circ$, 60$^\circ$) were also performed. In this way, exploiting the correlation between the residual energy measured by the device itself and the incident angle, it is possible to estimate the depletion depth of the SiC sensors at the FDV. 

\section{Validation of the energy loss tables of Ziegler}
\label{sec:validation}

The nominal total thickness of the SiC devices, obtained after the mechanical thinning procedure, is $t$ = 110 $\pm$ 1 $\mu$m, as confirmed by the micrometer measurement. This total thickness $t$ can be verified by measuring the residual energy $E_r$ in the PIPS detector when protons with $E_0=$ 6.00 MeV cross the SiC device at different angles of incidence $\theta$ \cite{knoll} by the following relation:
\begin{equation} \label{eq:e_r}
    E_r = E_0-\Delta E = E_0 - \int_0^{\frac{t}{\cos{\theta}}} \frac{dE}{dx} dx
\end{equation}
where $\Delta E$ is the energy loss in the SiC material, and $\frac{dE}{dx}$ is the ionization profile of 6.00 MeV protons in the same material (as shown in Figure \ref{fig:ranges}), according to the Ziegler tables \cite{Ziegler1985}. The first goal of the present work is the validation of these energy-loss tables. To empirically correct possible discrepancies between the nominal total thickness $t$ and its estimate from the residual energy measurement, the following normalization factor $k$ is introduced:
\begin{equation}
    k = \frac{1}{N}\sum_{i=1}^N\frac{E_0 - E_{r_i}}{\int_{0}^{t_i}{\frac{dE}{dx}dx}}
\end{equation}
where $N$ is the number of the measurements at the different angles $\theta_i$, and $t_i$ is defined as $\frac{t}{\cos{\theta_i}}$. The $k$ values obtained for each SiC device are listed in Table \ref{tab:k_values}. The obtained normalization factors correspond to a deviation smaller than 5\% and are in agreement among the different devices. From the weighted average of $k$ values, an overall normalization factor $K$ = 0.965 $\pm$ 0.001 is obtained. In this analysis, the energy loss in the Al ($\approx$ 1 $\mu$m) and Ni$_2$Si ($\approx$ 100 nm) layers were assumed to be negligible. 

\begin{table}
    \centering
    \caption{Values of normalization factors $k$ for the energy loss tables of Ziegler \cite{Ziegler1985} as deduced for 6.00 MeV protons on SiC material (see text), evaluated from the measurements on each SiC detector under study.}
\label{tab:k_values}
    \small
    \begin{tabular}{cccc} 
    \hline
    Wafer & Sensor ID & k \\
    \hline
    TT0012-11 & A41 & 0.960 $\pm$ 0.001\\
    TT0012-11 & A45 & 0.971 $\pm$ 0.001\\
    RA0089-27 & A106 & 0.959 $\pm$ 0.002\\
    RA0089-27 & A102 & 0.962 $\pm$ 0.003\\
    \hline
    \end{tabular}
\end{table}

\section{Measurement of the depletion depth}
\label{sec:depletion_depths}

\begin{figure}
    \centering
    \includegraphics[width=1\linewidth]{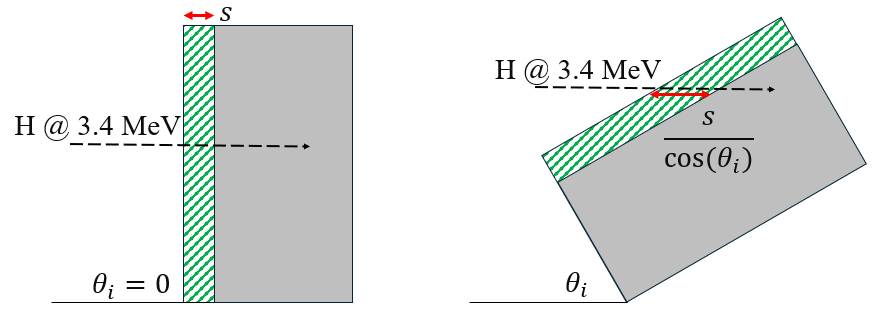}
    \caption{Scheme of the experimental setup for the 3.40 MeV proton irradiation from the SiC back side at different angles. The SiC active region and the dead layer are indicated as the gray and green hatched bands, respectively.}
    \label{fig:scheme}
\end{figure}

First characterizations of the SiC sensors suggested that the full depletion depth is less than 100 $\mu$m \cite{carbone2024}, depending on the doping concentration. For this reason, it is important to accurately determine the thickness of the dead layer on the back of the sensors, which could be larger than the expected 10 $\mu$m. This analysis was carried out by irradiating the SiC sensors with $E_0$ = 3.40 MeV protons from the back side (see Figure \ref{fig:scheme}). Increasing the incident angle $\theta_i$ of the protons, the pulse height of the SiC detector $E_{r_i}$ is reduced, due to the increase in ion path length $s$ in the dead layer. The thickness of this layer was deduced minimizing the $\delta(s)$ deviations, as defined in the following equation:
\begin{equation} \label{eq:e_r}
    \delta(s) = E_{r_i} - \left(E_{0_i}  - K \int_0^{\frac{s}{\cos{\theta_i}}} \frac{dE}{dx} dx\right)
\end{equation}
Different initial proton energies $E_{0_i}$ for the different incident angles were adopted to take into account the energy loss in the Al layer ($\approx$ 1 $\mu$m) on the back side of the SiC sensor . The $\frac{dE}{dx}$ ionization profile for 3.40 MeV protons was corrected adopting the $K$ factor deduced in Section \ref{sec:validation}. The dead layer values obtained from the average of the different incident angles $\theta_i$ for each SiC detector are listed in Table \ref{tab:deadlayer}. In order to compare the present results with those of Ref. \cite{carbone2024}, obtained using irradiation with a $^{228}$Th $\alpha$-source, we assumed the same $K$ normalization factor for the ionization profiles, thus obtaining the values reported in Table \ref{tab:deadlayer}. They are in agreement with the present work, thus validating the assumption of the same normalization factor for light incident particles (H and $\alpha$ in the present case). From the weighted average of these dead layer estimations, we deduced the depletion depth for each SiC sensor, as listed in Table \ref{tab:deadlayer}. In the case of the TT0012-11 wafer, the depletion depth is very close to the nominal one. On the contrary, the sensors belonging to the RA0089-27 wafer display a smaller depletion depth than the expected 100 $\mu$m. This confirms the worse performance obtained for this wafer, probably due to the not ideal doping profile, as already discussed in Ref.~\cite{carbone2024}.

\section{Conclusions}
\label{sec:conclusions}

We performed a characterization of state-of-the-art large area single pad SiC detectors in terms of depletion depth reached at the FDV. Sensors were produced from two different wafers, with different doping concentration. The TT0012-11 wafer was doped with $\approx$ 9 $\times$ 10$^{13}$ atoms/cm$^3$, which corresponds to a FDV of $\approx$ 780 V. The results of the present paper confirm that sensors belonging to this wafer meet the expected production requirements. In particular, they reach a depletion depth of $\approx$ 100 $\mu$m. On the contrary, the devices from the RA0089-27 wafer, doped with $\approx$ 3 $\times$ 10$^{13}$ atoms/cm$^3$ with FDV $\approx$ 270 V, display worse performances. We found a full depletion depth of $\approx$ 94 $\mu$m, significantly smaller than the nominal epitaxial layer. 

In nuclear physics applications, it would be much more beneficial working with detectors characterised by a FDV as low as possible, especially when working in a low-pressure gas environment, to reduce the probability of discharges and electrical instabilities. This is the case of the NUMEN project, which is planned to use more than 700 detectors coupled with a low pressure gas-tracker \cite{cappuzzello2021TDR}. The results of the present paper confirm that further improvements in the technologies for low doping concentrations are needed. New SiC sensors with the desired uniform epitaxial growth are in production, thanks to a new generation of reactors recently developed for this purpose. Characterizations similar to those described in the present paper will be performed on them as soon as they will be available.

\section{Acknowledgements}
\label{sec:acknoledgements}

The authors acknowledge the financial support for transnational access to the RBI accelerator facility by the project EURO-Labs funded from the European Union's Horizon Europe Research and Innovation programme under Grant Agreement No. 101057511. This project received funding from the European Union "Next Generation EU" (PNRR M4 ‐ C2 – Inv. 1.1 - DD n. 104 del 02‐02‐2022 - PRIN 20227Z4HB8). 



\bibliographystyle{elsarticle-num-names}

\bibliography{biblio.bib}


\end{document}